\documentclass[letterpaper]{article} 
\usepackage{aaai25}  
\usepackage{times}  
\usepackage{helvet}  
\usepackage{courier}  
\usepackage[hyphens]{url}  
\usepackage{graphicx} 
\usepackage{natbib}  
\usepackage{caption} 
\usepackage{algorithm}
\usepackage{algorithmic}

\usepackage{booktabs}

\usepackage{xcolor}

\usepackage{newfloat}
\usepackage{listings}
\DeclareCaptionStyle{ruled}{labelfont=normalfont,labelsep=colon,strut=off} 
\floatstyle{ruled}
\newfloat{listing}{tb}{lst}{}
\floatname{listing}{Listing}
\nocopyright

\title{The Spread of Virtual Gifting in Live Streaming: The Case of Twitch}

\author {
    Ji Eun Kim\textsuperscript{\rm 1},
    Seura Ha\textsuperscript{\rm 1},
    Sangmi Kim\textsuperscript{\rm 1, \rm 2},
    Libby Hemphill\textsuperscript{\rm 1}
}
\affiliations {
    \textsuperscript{\rm 1}School of Information, University of Michigan\\
    \textsuperscript{\rm 2}School of Education and Human Development, University of Miami\\
    jieunkim@umich.edu, seura@umich.edu, sxk2113@miami.edu, libbyh@umich.edu
}

\usepackage{bibentry}

\begin{document}

\maketitle

\begin{abstract}
This paper examines how gifting spreads among viewers on Twitch, one of the largest live streaming platforms worldwide. Twitch users can give gift subscriptions to other viewers in the chat room, with the majority of gifters opting for community gifting, which is gifting to randomly selected viewers. We identify the random nature of gift-receiving in our data as a natural experiment setting. We investigate whether gift recipients pay it forward, considering various gift types that may either promote or deter the spread of gifting. Our findings reveal that Twitch viewers who receive gift subscriptions are generally more likely to pay it forward than non-recipients, and the positive impact of gift-receiving becomes stronger when the recipient is the sole beneficiary of the giver’s gifting behavior. However, we found that gifts from frequent gifters discourage recipients from paying it forward, and gifts from anonymous gifters do not influence the likelihood of viewers becoming future gifters. This research contributes to the existing literature on the spread of online prosocial behavior by providing robust evidence and suggests practical strategies for promoting online gifting.
\end{abstract}

%

\section{Introduction}

Gifting has been studied as a fundamental component of human society across various contexts and disciplines. In the digital age, gifting has evolved with the development of information and communication technologies (ICTs). Modern gifting and donation systems, incorporating social media features, enable researchers to investigate large-scale online gifting and donation patterns. Understanding how people exchange gifts and make donations online is crucial for establishing and maintaining large-scale, active gifting systems, including crowdfunding platforms \cite{sisco2019examining}, live streaming services \cite{chaudhry2024express}, and gift exchange systems on social media \cite{kizilcec2018social} and online gaming environments \cite{bisberg2022gift}.

\begin{figure}[t]
\centering
\includegraphics[width=0.9\columnwidth]{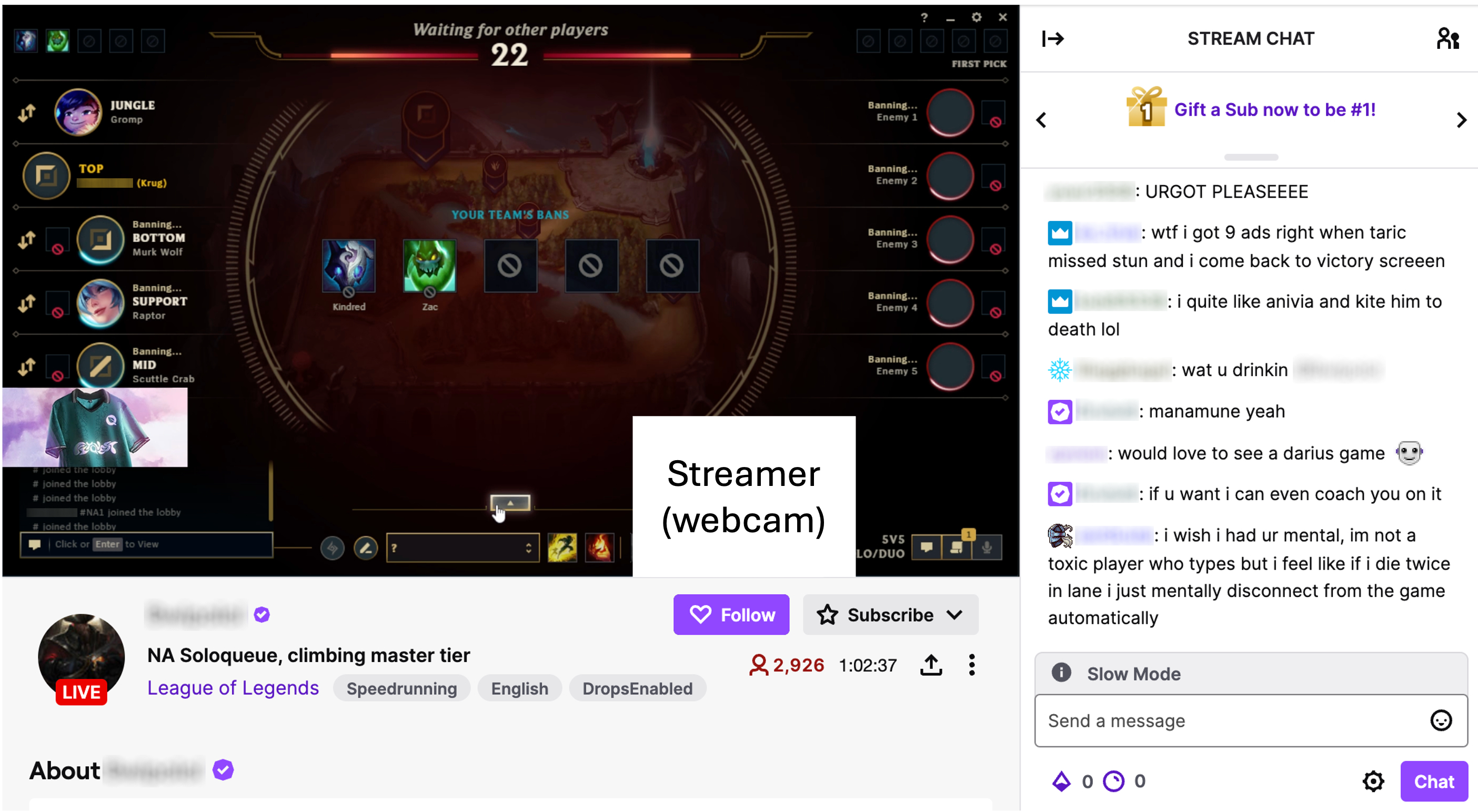}
\caption{Example Twitch stream interface. The streamer broadcasts video game content to the audience in real time, while viewers interact with each other through a chat window attached to the live stream. For privacy, the streamer and chatter nicknames have been removed from the screenshot.}
\label{fig1}
\end{figure}

This paper empirically investigates the spread of gifting in live streaming, focusing on Twitch. Live streaming has become a global phenomenon in online entertainment, offering new forms of user participation through real-time communication. Researchers have increasingly turned their attention to live streaming due to its extensive viewership and significant impact, particularly in the video game industry \cite{johnson2019impacts}. We selected Twitch for this research due to its vast global audience and the accessibility of its data. Twitch viewers can use a chat window attached to the video stream to post messages and give virtual gifts to streamers and other viewers in real time (see Figure \ref{fig1}). To date, there has been little quantitative analysis of gifting behavior within live streaming. We analyze a dataset of interactions from a random sample of Twitch streams to determine whether gifting is contagious among viewers and to investigate the conditions under which recipients choose to pay it forward. 

On Twitch, there are three main types of human users: streamers, viewers, and moderators. Many streamers also utilize automated bots to manage the chat room. It has been found that the majority of the audience consists of ``lurkers,'' viewers who watch the stream without participating in the chat \cite{wohn2020audience}. In this study, we use the term ``chatters'' to refer to users who participate in the chat. 

There are three main types of viewer engagement on Twitch: commenting, donating, and gifting. First, viewers can post text messages and emotes to communicate with the streamer and other viewers through a chat window attached to the video stream. Since commenting is free, it is the most basic form of user engagement.

Second, Twitch viewers can financially support streamers in various ways, such as subscribing to channels, purchasing memberships, or sending virtual currency donations. While viewers are not required to donate—since they can watch live streams and chat for free—some choose to give monetary rewards to streamers in appreciation of their entertainment efforts. Viewers can purchase the platform’s virtual currency, called Bits, at an exchange rate of approximately 100 Bits per USD and post messages along with the number of Bits they wish to donate.

\begin{figure}[t]
\centering
\includegraphics[width=0.9\columnwidth]{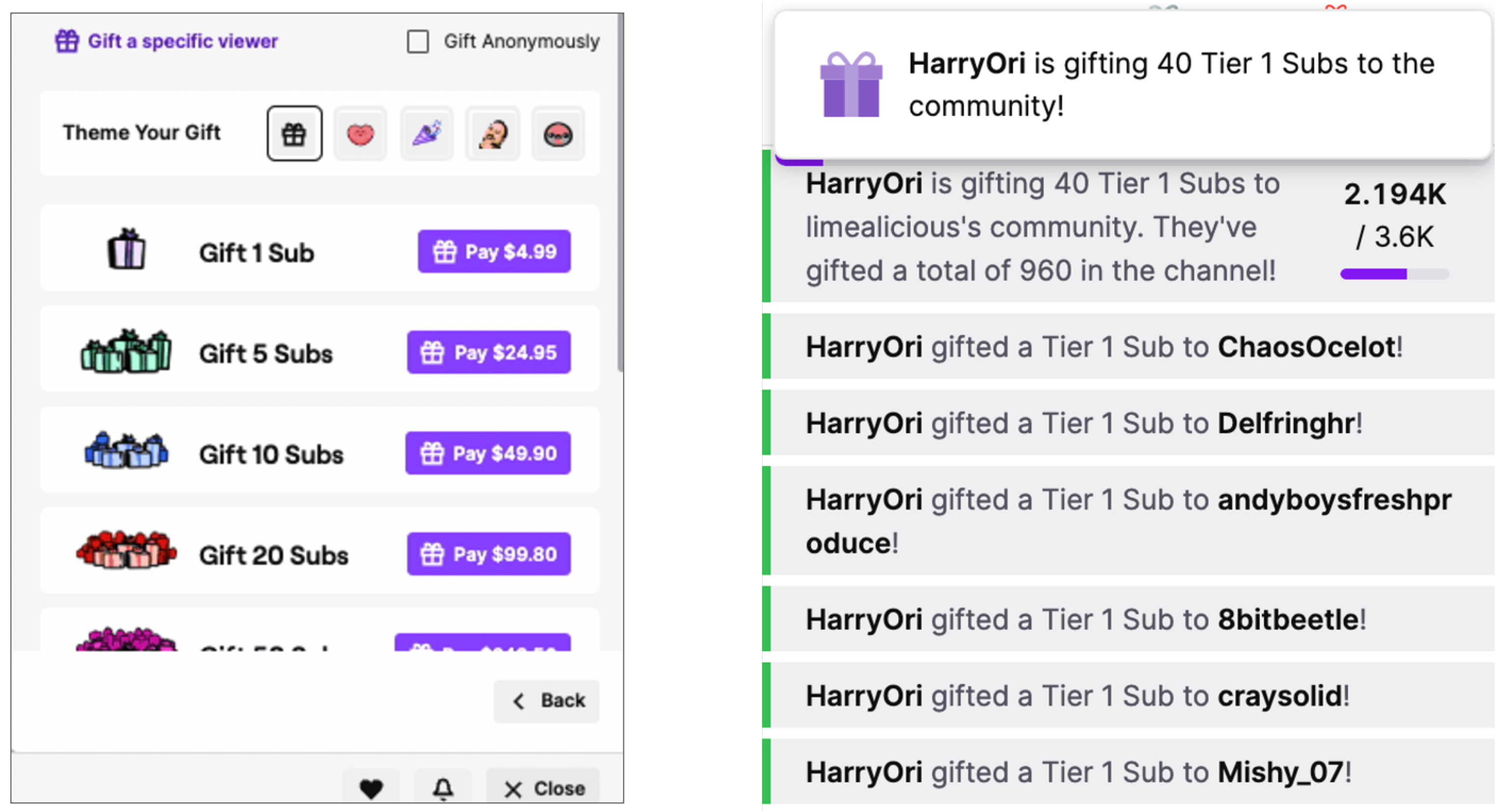}
\caption{The Twitch gifting feature allows users to gift a subscription to a specific viewer or to randomly selected viewers in the stream (Left). The announcement messages reveal the gifter and the recipients, noting that the user gave 40 gift subscriptions to the community. Additionally, it displays the total number of subscriptions gifted by the gifter to the channel, which is 960 in this example (Right).}
\label{fig2}
\end{figure}

Although Bits are intended solely for streamers, Twitch users can send virtual gifts, called gift subscriptions, to other viewers. Recipients of gift subscriptions enjoy several perks, such as access to special emotes and the ability to skip advertisements during the live stream. Users can gift a specific viewer in the chat room or randomly gift multiple subscriptions to non-subscribers in the chat room (see Figure \ref{fig2}\footnote{The images were retrieved from the Twitch website \cite{twitch_d, twitch_c}.}). Some gifters choose to remain anonymous. Throughout this paper, we use the term ``gift'' to refer to a gift subscription given by one viewer to another in the chat room, while ``donation'' refers to a monetary reward given by a viewer to a streamer.

To examine the conditions that encourage or discourage the willingness to gift, we analyze a dataset from 8,068 Twitch streams. Many Twitch users give gift subscriptions to randomly selected viewers in the chat room, and we leverage this random assignment of community gift subscriptions to conduct a causal inference analysis. Using mixed-effects logistic regression models, we estimate the effects of several gift-related features on the spread of gifting. Overall, we found that gift recipients are more likely to pay it forward than non-recipients, and the positive impact of gift-receiving on the spread of gifting becomes stronger when the recipient is the sole beneficiary of the gifting event. However, receiving a gift from a frequent gifter discourages future gifting by the recipient.

Our work offers both theoretical and practical contributions. First, our findings provide empirical evidence on the spread of online prosocial behavior, supporting social science theories regarding the factors that initiate and sustain a chain of goodwill. Second, the insights from this research can be applied to broader contexts of online gifting beyond live streaming. They can inform stakeholders aiming to promote prosocial behavior by proposing strategies, such as adjusting features related to the level of attention beneficiaries receive or the visibility of gifting activities.

\section{Related Work}

\subsection{Gifting and Contagion of Prosocial Behavior}

Gifting is a form of prosocial behavior defined as “all kinds of actions that benefit others, often at a personal cost to the actor \cite{thielmann2020personality}.” Because voluntary gifting incurs social and monetary costs for the giver, research has explored the motivations behind it. These motivations include relationship building and maintenance \cite{ruth1999gift, sherry1983gift}, symbolic communications that signal social, economic, and educational status \cite{camerer1988gifts, schwartz1967social}, self-representations \cite{batson2003altruism}, and pure hedonic and eudaimonic values such as pleasure and self-fulfillment \cite{anik2009feeling, lai2020good, wolfinbarger1993three}.

Although less attention has been given to the reactions of gift recipients compared to gifters and their motives \cite{lampinen2013indebtedness}, understanding these reactions is crucial for explaining how prosocial behaviors spread. Previous studies have found that people who receive benefits from others are likely to pay them back (i.e., to the giver) or forward (i.e., on to others) \cite{tsvetkova2014social, skaageby2010gift}. Recipients’ subsequent prosocial acts are often driven by two emotional reactions: feelings of indebtedness and gratitude \cite{peng2018reconsidering, schaumberg2009differentiating}. Beneficiaries of prosocial behavior often feel obliged to reciprocate because they have received support from others and may experience unhappiness or anxiety if they feel their reciprocation is insufficient \cite{nahum2011social}. Additionally, a separate line of research has shown that gratitude promotes prosocial behaviors, including paying it forward by helping others \cite{you2006gratitude, nowak2007upstream}. 

\subsection{Generalized Reciprocity in Online Communities}

While fundamental ideas about prosocial behaviors have long been developed and discussed within traditional, offline communities, the literature has recently expanded to encompass online communities. Online communities offer unique settings that facilitate the offering, obtaining, and witnessing of gift exchanges due to features that support interactions among community members \cite{bisberg2022gift, kwon2017s}. Although some perspectives suggest that gifting in online environments is less intimate and requires less effort compared to offline contexts—thereby potentially reducing the perceived value of gifts \cite{kwon2017s, shmargad2016online}—communication features such as messaging, posting, commenting, and tagging, as well as tools designed for gift exchange, enable immediate requests and distribution of gifts.

While these features allow individuals in online communities to give, receive, and reciprocate easily, research has found that people perceive a reduced need to engage in prosocial behavior in large online spaces compared to small offline communities \cite{tsvetkova2015contagion}. In online communities, prosocial behaviors are often based on the expectation of generalized reciprocity, which involves paying it forward rather than directly back to the giver \cite{lampinen2013indebtedness}. This suggests that recipients may choose to hide, exploiting online environments where social cohesion is weak, users are anonymous, and punishing defectors is difficult \cite{dubreuil2008strong}. Consequently, there remains significant uncertainty about the spread of online prosocial behavior, underscoring the need to investigate the mechanisms and conditions that encourage or discourage users from propagating such behavior.

\subsection{Gifting in Live Streaming}

Live streaming provides a large-scale and dynamic environment, allowing researchers to observe social patterns and behavioral contagion in interpersonal interactions among users, including gifting. Previous research primarily focused on the dynamics of gifting from viewers to streamers. One line of research explores a set of conditions that encourage donations to streamers. It has been found that user engagement is positively associated with donations to streamers \cite{yu2018impact}. Other studies have examined how streamer characteristics influence donation behaviors. For instance, streamer traits such as trustworthiness and attractiveness foster emotional attachment, which subsequently impacts viewers’ donation decisions \cite{li2021drives}. Another project also discovered that streamer characteristics, particularly gender, influence user engagement and donations on Twitch \cite{wolff2024audience}. They found that moderation significantly boosts user engagement and donations, with this effect being much stronger for female streamers.

While the aforementioned studies emphasize monetary engagement within streamer-viewer relationships, there is limited research on gifting among peer viewers in live streaming. Chaudhry, Wang, and Ouyang (2024) found that Twitch users who receive a gift subscription tend to reciprocate by giving other users subscriptions. However, they concluded that gifting among peer viewers did not lead to increased donations to streamers because, after receiving gifts from others, viewers feel more indebted to their peers than to the streamers. This finding can also be explained by the fact that people tend to exchange similar types of resources \cite{plickert2007s}.

\subsection{Research Gaps and Questions}

Based on the literature review, we identified several gaps for investigation. While various factors at the individual, social, cultural, and community levels influence pay-it-forward behaviors, there is a lack of comprehensive understanding of the circumstances under which gift recipients in online environments pay it forward. This gap makes it challenging to translate and expand the knowledge into diverse online contexts. Additionally, live streaming services, a relatively novel form of online community, have been studied from the gifter's perspective, usually employing subjective reports like surveys. Examining behaviors through observational data is critical to expanding our understanding of prosocial reciprocity. Moreover, prior research has mainly focused on the social interactions between streamers and their audiences rather than on interactions among viewers. The mechanisms underpinning gifting among viewers require further exploration. 

To address these gaps, this study examines whether gift recipients pay it forward within the live streaming environment and investigates specific factors that may influence the spread of gifting. These factors not only highlight the utility of social and donation features—such as selecting random recipients and revealing or concealing the identity and history of gifters in online communities—but have also been examined in other contexts as contributing or detrimental factors for generalized reciprocity. In the following paragraphs, we explain the factors that may influence gifting behaviors.

\subsubsection{Anonymous Gifters.} Gifting is often seen as a purposeful activity aimed at building social relationships or reputations \cite{berman2022prosocial}, which suggests that gifters usually prefer to be recognized. For instance, on GoFundMe, only about 20\% of gifts are given anonymously \cite{sisco2019examining}. Because it does not carry social capital benefits or advertise givers prosocial behaviors, anonymous gifting is typically viewed as purely altruistic \cite{siem2018attribution, tsang2019four}. Previous research has shown that receiving support perceived as unconditional and genuine makes recipients feel more grateful and motivates them to pay it back or forward \cite{ma2014gratefully}. Therefore, it is plausible to predict that gifts from anonymous gifters might enhance gratitude, as recipients are more likely to perceive them as unconditional and altruistic, thereby increasing the likelihood of recipients paying it forward. 

\subsubsection{Number of Beneficiaries.} Previous studies have examined how helping a single beneficiary versus multiple people at once affects the likelihood of beneficiaries paying it forward. Researchers suggest that beneficiaries are more likely to reciprocate when they are the sole recipients of prosocial behavior. In such cases, individuals tend to feel more grateful due to receiving special attention from the giver and feel more obligated to reciprocate, as they are more visible and less able to blend into the group \cite{kolyesnikova2008effects}. Additionally, recipients of individual gifts often perceive these gifts as larger than those given to a group, making them more inclined to reciprocate \cite{tsang2021special}. Recipients of group gifts are less likely to exhibit prosocial behavior, in part because the responsibility to pay it forward is diffused among all recipients \cite{tsvetkova2015contagion}. 

\subsubsection{Gifting Histories of Gifters.} When community gifting occurs in the chat room, Twitch displays the cumulative number of gifts given by the gifter as an announcement (see Figure \ref{fig2}). These gifting histories indirectly reveal the status of gifters, reflecting their financial commitment to the channel. Researchers have found that individuals often imitate high-status donors to associate themselves with a higher social rank \cite{kumru2010effect}. Another research reported imitation effects in Twitch chat rooms, where users mimic the commenting behaviors, such as spam messages, questions, and smile emotes, of high-status users \cite{seering2017shaping}. However, evidence from a live streaming platform in China suggests that viewers tend to send fewer and cheaper gifts when others are sending expensive gifts, feeling unable to compete monetarily in the chat room \cite{luo2024gift}. As a result, the influence of gifter status on a recipient's subsequent gifting behavior remains unclear.

We suggest that the factors discussed above are important determinants of a gift recipient's decision of whether to become a gifter and aim to address the following research questions:

\begin{itemize}
  \item \textbf{RQ1}: Do gift recipients pay it forward by giving gift subscriptions to other viewers?
  \item \textbf{RQ2}: Do gift recipients pay it forward after receiving gifts from anonymous gifters?
  \item \textbf{RQ3}: Are gift recipients more likely to pay it forward after receiving individual gifts instead of group gifts?
  \item \textbf{RQ4}: How do gifts from infrequent and frequent gifters affect the recipient's future gifting behavior?
\end{itemize}

\section{Data and Methods}

\subsection{Data}

We used the Twitch API\footnote{https://dev.twitch.tv/docs/api/} to collect user data, stream data, and chat logs for our study. Initially, we collected a random sample of 18,657 English-language streams, publicly broadcasted during the first two weeks of September 2022. Previous research has shown that only a few successful streamers attract large viewership, while most streamers struggle to gain an audience \cite{deng2015behind}. Our dataset confirmed this pattern, with approximately 18\% of streams containing no chat messages and 55\% containing more than 50 messages. Given that our research focused on gifting patterns involving many Twitch users, we excluded streams with few messages. The resulting dataset includes 8,068 streams, each with at least 10 unique users and between 50 and 10,000 chat messages. In total, the dataset includes 956,328 unique users and 11,272,948 chat messages (see Table \ref{table1} for descriptive statistics). Our dataset does not contain any personally identifiable information, and we have presented the results aggregated at the stream level.

\begin{table}[t]
\centering
\begin{tabular}{lrr}
    \toprule
    User type & Number of users & Number of messages \\
    \midrule
    Chatters & 916,139 (95.80\%) & 9,020,657 (80.02\%)\\
    Moderators & 33,591 (3.51\%) & 1,498,147 (13.29\%) \\
    Bots & 717 (0.07\%) & 638,053 (5.66\%) \\
    Streamers & 5,881 (0.61\%) & 116,091 (1.03\%) \\
    \midrule
    Total & 956,328 (100\%) & 11,272,948 (100\%) \\
    \bottomrule
\end{tabular}
\caption{Dataset overview.}
\label{table1}
\end{table}

\subsubsection{Bot Detection.} Most Twitch streamers rely on human moderators and/or bots to manage their chat rooms and engage with viewers. The Twitch API marks all messages posted by moderators in the chat logs; however, it does not differentiate between messages posted by human moderators and those posted by bots. Human moderators and bots show different activity patterns on Twitch. Human moderators customize their messages freely and engage with the chat at their discretion, while bots typically repeat predefined messages at regular intervals. For example, bots may remind viewers about upcoming events or prompt streamers to stay hydrated at set times. Additionally, bots react to specific situations; for instance, they express gratitude to viewers who donate money to the streamer. They can post multiple automated messages rapidly, often with better grammar. Many bots include the substring ``bot'' in their usernames and are present in many streams; for instance, 50.72\% of the streams in our sample used the StreamElements bot. In contrast, human moderators cannot be present in multiple streams due to physical limitations. To accurately identify bots in our dataset, we used the website that lists Twitch bots\footnote{https://twitchbots.info/}, finding 201 bots initially. We manually verified the accuracy of the list and identified several bot characteristics. Given that the website relies on crowd-sourced reports, some bots may remain undetected. We used the aforementioned criteria distinguishing bots from human moderators to identify these undetected bots, resulting in a final count of 717 unique bots. 

\subsubsection{Descriptive Statistics.} Out of the 8,068 streams, 47.15\% (3,804 streams) contained at least one gifting event. As illustrated in Figure \ref{fig2}, an announcement is made in the chat window immediately after a user gifts subscriptions to others. This announcement includes details such as the total number of gifts given to the channel by the gifter and the number of gifts given in a single event. In our dataset, a total of 65,732 gift subscriptions were distributed across 15,653 gifting events. Although 10.44\% (6,863) were directed to specific viewers by gifters, the vast majority—87.17\% of the gift subscriptions (57,301)—were randomly given to non-subscribing viewers. Only 2.39\% (1,568) were sent by anonymous gifters. Individual and group gifts account for 9.29\% (6,106) and 90.71\% (59,626) of the total number of gift subscriptions, respectively. The dataset includes 11,898 unique gifters and 64,691 unique recipients. Among streams with at least one gifting event, the median number of gift subscriptions given is 7.

\subsection{Natural Experiment Design and Randomization}

Gifting on Twitch offers a unique opportunity to study behavioral contagion through a natural experiment. Viewers have two options for gifting on Twitch: they can send gifts to randomly selected or specified viewers. When distributing gift subscriptions randomly, the platform uses an algorithm that prioritizes non-subscribing viewers in the chat room, followed by followers, moderators, and so on, while avoiding trolling users \cite{twitch_c}. We aim to examine if gift-receiving influences the recipient's subsequent gifting behavior within a causal inference framework. Because the majority of gifts were randomly distributed, these gifting events simulate a random assignment of treatment. In this framework, viewers who receive randomly distributed gifts can be considered \textit{treated} users, while other viewers, who do not receive gifts but are comparable to the treated users, serve as \textit{control} users. By “comparable,” we mean that the control users can serve as a counterfactual—representing what might have happened to the treated users if they had not been randomly selected to receive gifts. For the control group, therefore, it is important to select users from the same context, which we define as a \textit{gifting event}, to ensure meaningful comparisons.

\begin{figure*}[t]
\centering
\includegraphics[width=0.8\textwidth]{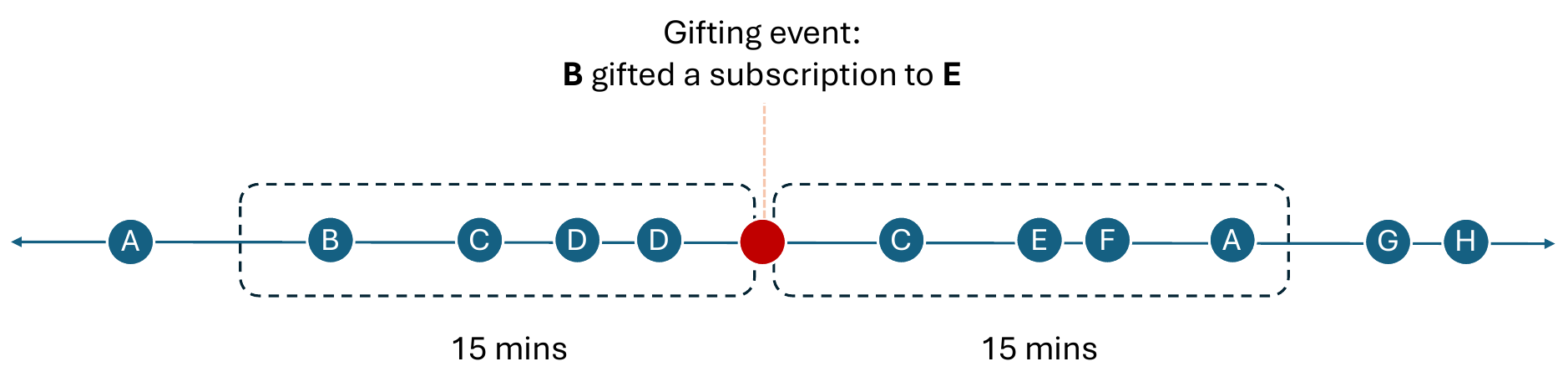}
\caption{Identification of a treated user and selection of control users. Each circle represents a message, with the letter inside indicating a viewer. In this example, the treated user is E, who received a gift subscription, while the control users are A, C, D, and F. Both treated and control users were non-subscribers and active in the chat within the specified timeframe.}
\label{fig3}
\end{figure*}

Figure \ref{fig3} illustrates the process of selecting treated and control users from a sample \textit{gifting event}. In this event, user E is in the treatment group because E received a gift subscription from user B. We used a 15-minute window to select control users, excluding gifter B. In other words, non-recipients who participated in the chat within a 15-minute window before and after the event were considered control users. We then excluded control users who were already subscribers to avoid inflating the control group’s gifting statistics and to ensure a fair comparison between the two groups. This resulted in non-subscribing users A, C, D, and F being identified as control users. Due to the lack of data on lurkers who do not leave any trace, our analysis included only recipients and non-recipients who posted at least one message in the chat room.

Previous research identified random gifting on Twitch as an exogenous shock but raised concerns about potential endogeneity issues due to Twitch's algorithm for assigning random recipients \cite{chaudhry2024express}. To address these concerns, Chaudhry, Wang, and Ouyang (2024) restricted their sample by excluding bots and users who had already subscribed before the gifting event. Similarly, we focused only on non-subscribing viewers, excluding streamers, moderators, and bots. The resulting dataset includes 4,127 gifting events across 1,808 streams. The numbers of users in the treatment and control groups are 6,130 and 255,943, respectively.

Our treatment and control groups were statistically similar before the gifting event. We measured the number of messages, the proportion of toxic commenters, and the proportion of donors who donated to streamers to assess the similarity between the two groups. When calculating the proportion of toxic commenters, we used toxicity scores of the messages as assessed by the Perspective API\footnote{https://perspectiveapi.com/}. We classified a commenter as toxic if their average comment toxicity score before the gifting event was 0.5 or higher, following the threshold set by previous research \cite{ghosh2021detecting}. As shown in Table \ref{table2}, the covariates are reasonably well-balanced between the treatment and control groups. This balance suggests that the random allocation in the community gifting system effectively achieves comparability between the two groups.

\begin{table*}[t]
\centering
\setlength{\tabcolsep}{8pt}
\renewcommand{\arraystretch}{1.1} 
\begin{tabular}{lllcc}
    \toprule
    Variable & \textbf{Treatment} & \textbf{Control} & \textit{t}-statistic & \textit{p}-value \\
     & Mean (SD) & Mean (SD) &  & \\
    \midrule
    Number of comments & 3.840 (10.56) & 3.612 (10.01) & 1.676 & 0.094 \\
    Proportion of toxic commenters & 0.005 (0.07) & 0.007 (0.08) & -1.748 & 0.081 \\
    Proportion of donors who donated to streamers & 0.005 (0.07) & 0.005 (0.07) & -0.075 & 0.940 \\
    \midrule
    \textit{N} & 6,130 & 255,943 & &  \\
    \bottomrule
\end{tabular}
\caption{Balance table.}
\label{table2}
\end{table*}

\subsection{Variables}

To answer our research questions, we performed mixed-effects logistic regression analyses with clustered standard errors at the stream level. We also included stream-level fixed effects to address unobserved heterogeneity across streams. The outcome variable is binary, indicating whether an individual becomes a gifter in any stream within the dataset after the gifting event. We aggregated comment-level data to the user level and included several binary variables as predictors in our models. The \textit{Gift-receiving} variable indicates whether a user is a gift recipient (1) or not (0). The \textit{Anonymous gifter} variable represents whether the gifter is anonymous (1) or not (0). The \textit{Individual gift} variable indicates whether the gift is given to a single recipient (1) or is part of multiple gifts given in bulk to two or more recipients during a gifting event (0). Additionally, we identified infrequent and frequent gifters based on the cumulative number of gifts they have given to the community. An \textit{Infrequent gifter} is defined as a gifter who falls below the 20th percentile of all gifters in the dataset, while a \textit{Frequent gifter} is someone above the 80th percentile. 

\section{Results}

We document results from four different model specifications. Model 1 includes our primary predictor, \textit{Gift-receiving}, as the sole independent variable. Model 2 incorporates the \textit{Anonymous gifter} indicator and its interaction term with \textit{Gift-receiving}. The interaction term separates the coefficient for \textit{Gift-receiving} term from Model 1 to examine the role of anonymity in pay-it-forward behavior. Specifically, the coefficient for the interaction term, \textit{Gift-receiving × Anonymous gifter}, indicates how the effect of receiving a gift changes when it is from an anonymous gifter, compared to a gift from an identified gifter, represented by the coefficient for the \textit{Gift-receiving} term in Model 2. Model 3 includes the \textit{Individual gift} variable and its interaction with \textit{Gift-receiving} to examine the effect of gift-receiving on paying it forward when it is given individually versus as part of a bulk gifting event. Finally, Model 4 includes the type of gifter (\textit{Infrequent Gifter} and \textit{Frequent gifter}) and its interactions with \textit{Gift-receiving}. The coefficients for the interaction terms indicate how the effect of receiving a gift varies based on the type of gifter. Across all models, we included stream-level fixed effects but omitted the fixed effects coefficients from the table. Under all model specifications, we found that receiving a gift increased the likelihood of gifting in the future.

\begin{table*}[t]
\centering
\setlength{\tabcolsep}{8pt}
\begin{tabular}{lllll}
    \toprule
     &\textbf{Model 1}&\textbf{Model 2}&\textbf{Model 3}&\textbf{Model 4}\\
    \midrule
    Intercept &-5.917*** &-5.924***&-5.896***&-5.918***\\
  &(0.076) &(0.076)&(0.078)&(0.079)\\
    Gift-receiving &1.005***&1.007***&0.912***&1.089***\\
      &(0.107) &(0.109)&(0.118)&(0.123)\\
    \midrule
    Anonymous gifter & &0.155& & \\
      & &(0.159)& & \\
    Gift-receiving $\times$ Anonymous gifter & &-0.027& & \\
      & &(0.480)& & \\
      \midrule
    Individual gift & & & -0.114& \\
      & & & (0.086) & \\
    Gift-receiving $\times$ Individual gift & & &0.547*& \\
      & & & (0.258) & \\
      \midrule
    Infrequent gifter & & & &-0.121\\
      & & &  & (0.107) \\
    Gift-receiving $\times$ Infrequent gifter & & & & 0.267\\
      & & &  & (0.366) \\
    Frequent gifter & & & &0.074\\
      & & &  & (0.088) \\
    Gift-receiving $\times$ Frequent gifter & & & &-0.560*\\
      & & &  & (0.278) \\
    \midrule
    AIC & 15,263.24 & 15,266.29 & 15,262.61 & 15,264.75\\
    BIC & 15,294.67 & 15,318.68 &15,314.99 & 15,338.08\\
    \bottomrule
\end{tabular}
\caption{Mixed-effects logistic regression results. The number of observations is 262,073. AIC and BIC refer to the Akaike information criterion and Bayesian information criterion, respectively. $^{*}p<0.05$; $^{**}p<0.01$; $^{***}p<0.001$.}
\label{table3}
\end{table*}

\subsection{Gift Recipients Promote the Spread of Gifting}

Model 1 in Table \ref{table3} shows that gift recipients are significantly more likely to give gifts in the future (\textit{b} = 1.005, \textit{p} $<$ 0.001), creating a virtuous cycle of gifting among viewers. The corresponding odds ratio (2.732) indicates that Twitch viewers who received a gift are 2.732 times more likely to pay it forward compared to non-recipients. The intercept value, representing the log-odds for a non-recipient becoming a gifter, is -5.917, implying that non-recipients have very low odds of becoming gifters.

\subsection{Gifts from Anonymous Gifters Do Not Influence the Spread of Gifting}

Model 2 reveals that receiving a gift from an anonymous gifter does not have a statistically significant effect on the spread of gifting for either non-recipients (\textit{b} = 0.155, \textit{p} = 0.332) or recipients (\textit{b} = -0.027, \textit{p} = 0.956). 

\subsection{Individual Gifts Promote the Spread of Gifting}

We test the effect of receiving an individual gift on the spread of gifting. According to the interaction term in Model 3, individual gifts have a significant impact on recipients’ willingness to reciprocate (\textit{b} = 0.547, \textit{p} $<$ 0.05). The odds ratio (1.728) indicates that when a user’s status changes from non-recipient to recipient, in combination with receiving an individual gift, the odds of becoming a future gifter, compared to not paying it forward, increase by a factor of 1.728.

\subsection{Gifts from Frequent Gifters Discourage Recipients from Gifting in the Future}

To address our final question, we examine whether the likelihood of a recipient exhibiting pay-it-forward behavior differs depending on whether they received a gift from an infrequent gifter or a frequent gifter. Model 4 shows that gifts from infrequent gifters do not significantly affect either non-recipients (\textit{b} = -0.121, \textit{p} = 0.257) or recipients (\textit{b} = 0.267, \textit{p} = 0.466). 

While gifts from frequent gifters do not significantly affect non-recipients (\textit{b} = 0.074, \textit{p} = 0.402), they discourage recipients to pay it forward (\textit{b} = -0.560, \textit{p} $<$ 0.05). The odds ratio for the interaction (Gift-receiving $\times$ Frequent gifter) is 0.571, which indicates that as user status changes from non-recipient to recipient, along with receiving a gift from a frequent gifter, the change in the odds of becoming a future gifter compared to not paying it forward is 0.571. In other words, after receiving a gift from a frequent gifter, recipients are less likely than non-recipients to gift in the future.   

\section{Discussion}

\subsection{Main Findings}

Understanding how prosocial behavior like gifting spreads is crucial for promoting positive online environments for users. We aimed to provide causal explanations for the spread of gifting and to identify the conditions under which gifting is more likely to spread on Twitch. Overall, gift recipients are more likely to pay it forward than non-recipients, indicating that merely observing gifting behavior is not enough to trigger gifting. This finding aligns with prior research suggesting that beneficiaries of prosocial behavior are more likely to pay it forward, while merely observing others engage in such behavior does not promote its spread \cite{tsvetkova2014social}. While Chaudhry, Wang, and Ouyang (2024) found that receiving a gift increases the likelihood of paying it forward in music streams on Twitch, we suggest that this effect also applies to streams of different genres, as our study focused on video game streams.

We also estimated the effects of various gift-related features on the spread of gifting. Our results show that gifts from anonymous gifters do not contribute to the spread of gifting among Twitch viewers. One possible explanation is that anonymity may dilute the perception of the giver, leading recipients to feel less grateful. It is rare for Twitch gifters to remain anonymous, with only 2.39\% of the gifts in our sample being given by anonymous gifters. The low prevalence and negligible impact of anonymous gifting suggest that gifting on Twitch is predominantly driven by social factors such as signaling and status-seeking, as people generally seek recognition for their contributions. 

Next, Twitch viewers are more likely to pay it forward when they receive a gift as a sole beneficiary. This might be because they perceive individual gifts as more special than group gifts given to multiple users, leading to a greater sense of gratitude or obligation to reciprocate, even if the value and type of gift each recipient receives are identical in both scenarios. This aligns with previous findings that receiving gifts perceived as intentional and personal results in a greater sense of gratitude \cite{peng2018reconsidering}, which in turn leads to a tendency to pay it forward. Another possible explanation for the positive impact of individual gifts is that imitating individual gifting is easier and less costly than group gifting, as it involves just one beneficiary.

Lastly, we found that Twitch viewers are not influenced by gifts from infrequent gifters but are less likely to become gifters themselves if they receive gifts from frequent gifters. The negative influence of frequent gifters on the spread of gifting may be due to the diffusion of responsibility \cite{tsvetkova2015contagion}, where recipients believe that frequent gifters will continue to give gifts to others, reducing their own perceived obligation to gift. Another possible explanation is that recipients may forgo engaging in gifting because they cannot compete with a frequent gifter in terms of financial commitment to the community. Although individuals often attempt to imitate high-status users \cite{kumru2010effect}, status-seeking can be competitive in settings where reputation systems exist \cite{lampel2007role}. Previous research about the competitive nature of gifting in live streaming noted that users reduce the frequency and value of their donations to streamers as competition intensifies, perceiving greater spending as less meaningful \cite{luo2024gift}. Similarly, recipients may feel unable to reciprocate at the same level as givers who have already made significant financial contributions to the community, thus deciding not to attempt paying it forward.

\subsection{Implications}

The findings of this research have several theoretical implications and propose several research questions in need of further investigation. One of our key contributions is the focus on the spread of gifting among viewers and generalized reciprocity, rather than on the parasocial relationships and interactions between streamers and viewers. Our work also contributes to the literature by proposing a robust causal inference framework to study online prosocial behavior. Observing causality in prosocial behaviors is challenging outside of randomized controlled experiments. Even in controlled experiments, results often suffer from external validity issues unless thoroughly designed and executed. We utilized highly granular user behavior data from Twitch, capturing detailed logs of interactions and gifting events. By leveraging Twitch’s community gifting feature, we identified a natural experiment setting where gift-receiving occurs largely by chance. This approach allowed us to make meaningful comparisons between gift recipients and non-recipients. 

Our findings expand the current literature on gifting and pay-it-forward behaviors from the recipients’ perspectives, which have been underexplored compared to the givers’ perspectives. Specifically, we identified conditions under which gifting behavior spreads (i.e., when viewers receive individual gifts) and does not (i.e., when gifts come from frequent gifters). The insights obtained from this research extend our knowledge of what affects the willingness of recipients to engage in prosocial behavior and could be applied to explaining broader contexts of online gifting beyond live streaming. Additionally, our results highlight the importance of investigating the characteristics of virtual gifts, along with online interactions and environments (e.g., anonymity, a large number of simultaneous connections) that may encourage or discourage users from reciprocating, as opposed to offline settings.

This research also has practical implications. Based on our finding that receiving an individual gift increases future gifting, live streaming and other online gifting platforms might consider making recipients more conspicuous to garner more attention, which can in turn elicit more reactions including gifting from them. Another strategy could be to adjust the visibility of user gifting activities, as the presence of frequent gifters negatively affects recipients' engagement in gifting. However, we should be cautious about adopting this approach, as it risks losing the benefits of the reputation system. Frequent gifters might become demotivated and reduce gifting if their contributions are not visible and they cannot gain social reputation in the chat room. A deeper understanding of the mechanisms behind online gifting can enhance strategies for crowdfunding and donation platforms. Additionally, identifying ways to increase gifting could support the sustained growth and viability of the live streaming industry, as subscriptions are a major source of income for professional streamers \cite{johnson2019and}.

\subsection{Limitations and Future Work}

This study has several limitations. First, the Twitch API does not provide detailed information about streamers and their video content. Future research could include this information to provide a more comprehensive understanding of the relationship between stream-related features and gifting in streaming environments. Although we could not account for streamer-specific or content-related variables in our analysis, we attempted to mitigate the impact of these factors. We did this by including stream-level fixed effects and ensuring that both treated and control viewers were collected within a specific timeframe for each gifting event, thereby exposing them to the same content within the stream.

Second, our dataset lacks information on lurkers, as we included only viewers who commented at least once through chat logs. This limitation prevented us from testing the effect of gift-receiving on lurkers, as it was impossible to precisely identify viewers who were lurkers during each gifting event. Additionally, the observation period is not long enough to analyze long-term effects. Future research could model artificial lurkers to incorporate them into the analysis or assess the long-term effects of gift-receiving in online settings.

Another limitation is the generalizability of our findings, as live streaming is a unique setting that differs from other types of online environments. For instance, the purpose of community gifting on Twitch—which is the focus of this study—is less connected to relationship development and maintenance because gifts are randomly distributed and users remain anonymous. As a result, gifters expect only a minimal chance of being repaid or witnessing recipients’ reactions. However, givers can still be recognized as contributors to the channel and build a social reputation because their contributions are publicly announced in the chat room, as illustrated in Figure \ref{fig2}. These gifting-related features differ from those of gift exchanges on social media, which aim to build, maintain, and strengthen relationships among family or friends in social networks, rather than focusing on building a reputation among strangers. Future research could explore other types of online gifting features and their impacts on the spread of gifting. 

After receiving gifts, some recipients choose to pay it forward, as shown in this study, or express gratitude toward the gifters in the chat window. Future research could also examine how receiving a gratitude message from a recipient influences a gifter's willingness to increase gifting.

\section{Conclusion}

This study analyzes Twitch live streaming data to examine the effect of gift-receiving on a recipient’s subsequent gifting behavior. Our results reveal that Twitch viewers are more likely to pay it forward after receiving gifts. While individual gifts further encourage the spread of gifting, gifts from frequent gifters decrease recipients’ future gifting. Gifts from anonymous gifters do not affect the likelihood of viewers becoming future gifters.

\bibliography{aaai25}

\end{document}